\documentclass[12pt,manuscript]{aastex}

\slugcomment{Nov 20, 2006}

\newcommand{\kms}{km s$^{-1}$}

\shorttitle{The JCMT Spectral Legacy Survey}
\shortauthors{Plume et al.}

\begin{document}

\title{The JCMT Spectral Legacy Survey}

\author{R. Plume\altaffilmark{1}, G. A. Fuller\altaffilmark{2}, F.
  Helmich\altaffilmark{3, 20}, F. F. S. van der Tak\altaffilmark{3}, 
H. Roberts\altaffilmark{8}, 
J.  Bowey\altaffilmark{5}, 
J. Buckle\altaffilmark{4}, 
H. Butner\altaffilmark{15}, 
E. Caux\altaffilmark{17}, 
C. Ceccarelli\altaffilmark{16}, 
E. F.  van  Dishoeck\altaffilmark{14}, 
P. Friberg\altaffilmark{15}, 
A. G. Gibb\altaffilmark{13} 
J. Hatchell\altaffilmark{7}, 
M.  R. Hogerheijde\altaffilmark{14}, 
H. Matthews\altaffilmark{11}, 
T. J.  Millar\altaffilmark{8}, 
G.  Mitchell\altaffilmark{12}, 
T. J. T. Moore\altaffilmark{9}, 
V.  Ossenkopf\altaffilmark{21, 3, 20},
J. M. C. Rawlings\altaffilmark{5}, 
J. Richer\altaffilmark{4}, 
M. Roellig\altaffilmark{21}, 
P. Schilke\altaffilmark{18}, 
M. Spaans\altaffilmark{20} 
A. G. G. M. Tielens\altaffilmark{19}, 
M. A. Thompson\altaffilmark{6}, 
S.  Viti\altaffilmark{5}, 
B.  Weferling\altaffilmark{15}, 
Glenn J.  White\altaffilmark{10, 22} 
 J. Wouterloot\altaffilmark{15}, 
J.  Yates\altaffilmark{5}, 
M.  Zhu\altaffilmark{15, 11}}

\altaffiltext{1}{Dept. of Physics and Astronomy, University of Calgary, 2500 University Dr. NW, Calgary, AB, T2N 1N4, CA}
\altaffiltext{2}{School of Physics and Astronomy, University of Manchester, Sackville Street, PO Box 88, Manchester M60 1QD, UK}
\altaffiltext{3}{SRON Netherlands Institute for Space Research, PO Box 800, 9700 AV, Groningen, NL}
\altaffiltext{4}{Cavendish Laboratory J J Thomson Avenue Cambridge CB3 0HE, UK}
\altaffiltext{5}{Department of Physics and Astronomy, University College London, Gower St.,
London, WC1E 6BT, UK}
\altaffiltext{6}{Centre for Astrophysics Research, Science \& Technology Research Institute, University of Hertfordshire, College Lane, Hatfield, AL10 9AB, UK}
\altaffiltext{7}{University of Exeter, UK}
\altaffiltext{8}{School of Mathematics and Physics, Queen's University Belfast, Belfast BT7 1NN}
\altaffiltext{9}{Astrophysics Research Institute, Liverpool John Moores University, Twelve Quays House, Egerton Wharf, Birkenhead, CH41 1LD, UK}
\altaffiltext{10}{The Open University, Walton Hall, Milton Keynes MK7 6AA, UK}
\altaffiltext{11}{Herzberg Institute of Astrophysics/National Research Council of Canada, 5071 W. Saanich Rd, Victoria, BC, V9E 2E7, CA}
\altaffiltext{12}{St. Mary's University, Halifax, NS, CA}
\altaffiltext{13}{Department of Physics and Astronomy, University of British Columbia, 6224 Agricultural Road, Vancouver, BC, V6T 1Z1, CA}
\altaffiltext{14}{Leiden Observatory, Leiden University, PO Box 9513, 2300 RA, Leiden, NL}
\altaffiltext{15}{Joint Astronomy Centre, Hilo, HA, US}
\altaffiltext{16}{LAOG, Grenoble, FR}
\altaffiltext{17}{CESR-UPS/CNRS, BP 4346, 31028 Toulouse Cedex 4, FR}
\altaffiltext{18}{MPI, Bonn, GER}
\altaffiltext{19}{MS 245-3, NASA Ames Research Center Moffett Field, CA 94035-1000, US}
\altaffiltext{20}{Kapteyn Astronomical Institute, University of Groningen, P.O. Box
800, 9700 AV Groningen, NL}
\altaffiltext{21}{I. Physics Institute, University of Cologne, ZŸlpicher Str. 77, 50937 K\"oln, GER}
\altaffiltext{22}{CCLRC Rutherford Appleton Laboratory, Chilton, Didcot Oxfordshire, OX11 9DL, UK}


\newpage

\begin{abstract}

  Stars form in the densest, coldest, most quiescent regions of molecular
  clouds. Molecules provide the only probes which can reveal the dynamics,
  physics, chemistry and evolution of these regions, but our understanding of
  the molecular inventory of sources and how this is related to their physical
  state and evolution is rudimentary and incomplete. The Spectral Legacy
  Survey (SLS) is one of seven surveys recently approved by the JCMT Board.
    Starting in 2007, the SLS will produce a spectral imaging survey
  of the content and distribution of all the molecules detected in the 345 GHz
 atmospheric window (between  332 GHz and 373 GHz) towards a sample of 5 sources.  Our
  intended targets are: a low mass core (NGC1333 IRAS4), 3 high mass cores spanning a range 
  of star forming environments and evolutionary states
  (W49, AFGL2591, and IRAS20126), and a PDR (the Orion Bar).  The SLS will use
  the unique spectral imaging capabilities of HARP-B/ACSIS to study the
  molecular inventory and the physical structure of these  objects,
  which span different evolutionary stages and physical environments, to probe
  their evolution during the star formation process.  As its name suggests,
  the SLS will provide a lasting data legacy from the JCMT that is intended to
  benefit the entire astronomical community.  As such, the entire data set
  (including calibrated spectral datacubes, maps of molecular emission, line
  identifications, and calculations of the gas temperature and column density)
  will be publicly available.

\end{abstract}


\section{Introduction}
\label{sec:intro}

In early 2006, the James Clerk Maxwell Telescope (JCMT) on the summit of Mauna
Kea, Hawaii began installation of a new generation of scientific instruments
SCUBA-2  and HARP-B/ACSIS (Holland et al. 2006; Smith et al. 2003; Buckle et al. 2006), which will
revolutionize our understanding of the sub-millimetre sky.  HARP-B (Heterodyne
Array Receiver Programme) is a 16-element heterodyne array receiver (in a 4x4
pixel configuration with a 2$'$x2$'$ field of view) designed to operate between 325-375 GHz with a receiver
noise temperature of $< 330$~K (SSB) averaged across the array.  HARP-B was
built by MRAO at the University of Cambridge (UK) in collaboration with the
Joint Astronomy Centre (Hawaii), the Astronomy Technology Centre at the Royal
Observatory in Edinburgh (UK), and the Herzberg Institute of Astrophysics in
Victoria (CA).  The SIS junctions at the heart of HARP-B were provided by the
Delft Institute of Microelectronics and Submicron-technology (DIMES) in the
Netherlands.  HARP-B is designed to work with the JCMT's new spectrometer -
the ACSIS correlator (Auto-Correlation Spectrometer and Imaging System).
ACSIS was built by the Dominion Radio Astronomy Observatory in Penticton, BC
(CA) and the Astronomy Technology Centre at ROE (UK),and allows 16 simultaneous IF inputs with a 2 GHz bandwidth.  

While SCUBA-2 will allow astronomers to quickly map the continuum emission at
submillimeter wavelengths and so examine the nature and distribution of the
cold dust, HARP-B/ACSIS will allow them to efficiently map the molecular gas.
Molecular emission lines can provide direct information on the physical
conditions (density, temperature, pressure, and optical depth) and kinematics
of the emitting gas. Given the high speed and sensitivity of HARP-B/ACSIS,
combined with the largest aperture telescope at an excellent observing site,
there will be no comparable instrument for mapping large areas of the sky at
sub-millimetre wavelengths or for surveying molecular emission across the
entire 345 GHz band.

Large scale, well coordinated surveys were called for to fully exploit these
exciting new instruments and, in response, the JCMT Legacy Surveys were
born.  These surveys are ambitious, community-driven projects involving over
100 astronomers from Canada, the UK, the Netherlands and a few additional
non-partner countries. This paper describes the goals of the JCMT Spectral
Legacy Survey (SLS), one of the seven Legacy Surveys approved by the JCMT
Board of Directors following a rigorous peer-review process.  In a two-year
period starting in 2007, the SLS will receive 187 hours of Band 4 weather
conditions ($0.12 < \tau_{225 GHz} < 0.2$), in order to survey the molecular
emission of a variety of star forming regions.  The SLS
will provide a lasting data legacy from the JCMT that is intended to benefit
the entire astronomical community by fully exploiting the spectral imaging
capabilities of HARP-B.  The SLS is led by coordinators from the three partner countries: G. Fuller (principal investigator) for the UK (gary.a.fuller@manchester.ac.uk), R. Plume for Canada (plume@ism.ucalgary.ca), and F. Helmich for the Netherlands (f.p.helmich@sron.rug.nl).

\subsection{Scientific Background}
\label{sec:background}

It is well-known that stars form in molecular clouds, and most stars form in
clusters in Giant Molecular Cloud (GMC) cores. Although this statement is
commonly accepted, the details of the process are not well known. For both low
mass and high mass stars, several phases in the star formation process have
been recognized (e.g. Shu et al. 1987; McKee \& Tan 2003). These include, first, a
precollapse cloud-core phase characterized by low temperatures, second, a
deeply embedded phase where a cool, protostellar object is surrounded by warm
gas, an infalling envelope and stellar winds, and, third, later phases where
strong stellar winds have progressively cleared the immediate environment and
exposed the newly formed star and its surrounding protoplanetary disk.

Although complicated, it is possible to build a physical picture of evolving,
star-forming cores.  Since these cores are extended objects which also contain
small scale structures, we will require detailed observations sensitive to a range of size
scales and physical conditions.  Interferometers can provide the spatial
distribution of dust and gas at the smallest size scales while single dish
telescopes equiped with large format, wide bandwidth, detectors are capable
of simultaneously imaging large number of molecules (albeit at lower spatial
resolution).  With proper physical modelling, the combination of the two
lead to a better understanding of the physics of cores and, therefore, help to
constrain the formation mechanism of stars and stellar clusters. For example, Van der
Tak et al. (2000) have revealed the detailed physical structure of a dozen
regions of high-mass star formation including AFGL 2591 using a combination of
single-dish and interferometer observations, which made it possible to assign
different velocity and excitation components to different physical parts of the core.

The dynamical and physical evolution during star formation is also accompanied
by a strong chemical evolution. For example, the cold cloud-core phase is
dominated by relatively simple molecular species. In contrast, the warm gas
around protostellar objects is characterized by surprisingly complex species.
This chemical evolution points towards the interplay of a multitude of
chemical processes, including ion-molecule reactions at low temperatures in
dark cloud cores, accretion and reaction on cold dust grains, and evaporation
once the newly formed star turns on (e.g. van Dishoeck \& Blake 1998). While
the composition, origin, and evolution of interstellar molecules in regions of
star formation is an important topic in its own right, observations of
molecules can also provide direct information on the physical conditions of
the emitting gas and hence of the physical evolution of a star forming region.
The large number of rotational lines from low-lying levels of a species and
its isotopologues - each with its own collisional and radiative excitation rates
- probe the density, temperature, and optical depth of gas very effectively
over a wide range of conditions relevant for regions of star formation. Furthermore, these
molecules play an important role in regulating the temperature, and hence pressure,  via
the process of heating and cooling through line-absorption and emission.

To date about 151 different molecular species have been detected in
interstellar and circumstellar gas (CDMS\footnote{http://www.cdms.de}), but few of these species have been studied in
more than a few sources. Our understanding of the molecular inventory of
sources and its distribution is poor: few sources have been systematically
surveyed at any frequency band and most such surveys have been limited to a
single position toward a source. None to date have imaged every molecular
line (see Blake et al. 1996). However, regions of star formation are complex by nature. Stars rarely
form in isolation: protostars often form in binary systems which are
themselves in groups or clusters. Furthermore, star formation generally
involves simultaneous infall and outflow, circumstellar disks, as well as the
presence of regions of warm, dense gas (hot cores) near the protostar itself.
Each of these regions is characterized by its own density, temperature, and
velocity structure as well as its unique chemical composition. Fully
disentangling this spatial complexity requires mapping in a large number of
transitions of a variety of molecules.

The study of the formation of high-mass stars is especially complicated by the
fact that chemistry time-scales (and, therefore, the time scale for cooling
through line emission) are comparable to the accretion time scale. Also the
large density and temperature gradients involved complicate the physical
modelling. Therefore, detailed observations are key to understanding the
processes involved in creating the stellar clusters that shape the Galaxy.
Over the last decades, there have been a few studies of the chemical
composition of prominent star-forming regions like Orion (Leurini et al. 2006; Comito et al. 2005;
White et al. 2003;
Schilke et al. 1997, 2001;  Blake et al. 1987; Sutton et al 1991), Sgr B2
(Nummelin et al. 1998, Goicoechea et al. 2004) or W3 (Helmich \&
Van Dishoeck 1997), the most infrared bright  star-forming region in
the sky. However, only recently have spectral surveys of less luminous sources
become feasible. In general, being less luminous implies that the source is more deeply
embedded and thus, in an earlier evolutionary stage where outflows have not
yet cleared away the molecular environment.  For example, the results of the
spectral survey of IRAS16293-2422 (initiated by the results of Cazaux et al.
2003 and shown in its first results by Parise et al. 2005 on HDO) enable us to
probe the chemistry in an unprecedented way and extend our knowledge to
earlier stages of star-formation.

The
complete molecular inventory and thus, the physical, chemical and
evolutionary status of sources are only full accessible through
unbiased spectral surveys. The
factor of 16 increase in mapping speed provided by HARP-B together with the
spectral coverage and resolution of ACSIS provide the most potent instrument
available to obtain the molecular inventory in sources in the 345 GHz
band.

The 325 GHz to 373 GHz spectral range is rich in known molecular transitions: Lovas (2004) lists 866
transitions arising from 82 different molecular species. Many of the
transitions in this band have a range of excitation energies from tens to hundreds of Kelvin, making this window particularly useful for studying material
which is heated above the $\approx$ 10~K of ambient molecular gas in clouds, and to constrain
the physical and chemical structure on scales smaller than the beam.
However this band has not been thoroughly explored in a variety of different sources,
 even though it is particularly
important in the future of submillimetre astrophysics since it corresponds to
the ALMA frequency 7 band which the ALMA Design Reference Science Plan
suggests will be used for more than 30\% of the ALMA observing in all areas of
science.

\section{Science Goals}
\label{sec:goals}

The goal of the SLS is to use molecular line observations to help understand
the evolution of young stellar objects spanning a range of evolutionary stages
and environmental conditions. This will be achieved by obtaining a complete
molecular line inventory in three different types of sources: low mass
protostars, young high mass sources, and photon-dominated regions (PDRs).
Combining these data with detailed chemical and physical models of the
sources, the survey aims to improve our understanding of molecular tracers as
probes of the astrophysics and evolution of sources.  In particular, this
survey will:
 
 \begin{enumerate}
 \item Provide an inventory of the column densities and spatial distributions
   of the molecular species towards sources of different types and age.
 \item Provide a broad-based derivation of the physical and chemical conditions (i.e.
   temperature, density, velocity, and chemical structure) of the sources on scales both smaller and larger
   than that of the primary beam.
 \item Identify important chemical diagnostics of the physical and chemical
   processes occurring in different types of sources and different
   evolutionary states.
 \end{enumerate}
 
 Beyond this, the survey will provide a uniform dataset for future
 explorations of these sources to the general astronomical community.  In
 conjunction with similar, higher frequency, surveys being planned for the
 Herschel Space Observatory, it will also help us determine what constitutes a
 typical spectrum in sources with different environmental and evolutionary
 states.

\subsection{Molecular Inventory}
 
Emission features in each source will be identified using the JPL\footnote{http://spec.jpl.nasa.gov/}, Cologne\footnote{http://www.cdms.de} and
other available line catalogues.  Tools for this type of analysis (e.g.
CASSIS\footnote{http://pc-126.cesr.fr/} ; XCLASS - Schilke et al. 2006) are currently under
development. The spatial information and the different linewidths likely to be
seen at different positions will be used to aid in the identification of
blended or confused emission features. Unidentified features will be examined
to determine whether they are consistent with being undocumented transitions
of other species present in the spectra. The line velocity and frequency,
intensity, integrated intensity and width, will be determined initially
through Gaussian fitting and/or a 1D clumping analysis, and then catalogued.

The unique capability of HARP-B/ACSIS to simultaneously map sources spatially
and spectrally will be of key importance in disentangling the spatial
complexity of sources and unraveling the contribution of the different
components of the environments (outflows, infalling material, disk, quiescent
cloud). Specifically, we will use the images, linewidths and excitation of the
different species to identify the signatures of these different components so
as to be able to separately determine their physical properties. Column
densities will then be calculated from the line integrated intensities
initially using rotation diagrams or statistical equilibrium
modelling (Sch\"oier et al 2005). We will
derive column densities for the molecules with the brightest and greatest
number of transitions in the survey. Ultimately it will be a goal to calculate
the spatial abundance structure for the majority of the detected species, although this
will be more difficult and may require additional observations of lines at
other frequencies for molecules with relatively few transitions in the SLS
survey band.

Apart from the numerous lines we expect to identify through the SLS, we also
expect to find a number lines which we will be unable to identify.  These "U-lines" will
be flagged for future analysis.
 
\subsection{Physical Conditions \& Abundance Profiles}
 
Millimetre and submillimetre continuum observations will be used to determine
the temperature and density distribution for each source (after correcting the
continuum flux for contamination by the line emission). For some of the
sources such data already exist, for others these will be obtained during
other observations.  Initially the modelling will be one dimensional in order
to rapidly estimate the molecular column densities and distributions. However
one dimensional models are unlikely to adequately reproduce the observations
and so we plan to extend the modelling to two dimensions and, if possible, to 3 dimensions.
 
Molecular line observations can also provide information about the physical
conditions in these sources.  The most obvious of these is the use of CO to
trace the structure of the bulk of the gas in molecular clouds and the small
masses of gas at high velocities in outflows from young stars. Rotation
diagram analyses of various molecules will also be
used to determine the kinetic temperatures and column density of the gas.  Furthermore,
for molecules that have known collision rates,  Large Velocity Gradient codes can also be used
to determine the gas densities, column densities and temperatures.
 
With the density and temperature structure determined, the emission expected
for different abundances and distributions of the molecular species will be
modelled using radiative transfer models such as RATRAN (Hogerheijde \& van
der Tak 2000), to determine which best reproduce the observed spectra.
Initially this will be done adopting trial abundance profiles assuming
constant abundance, jump, anti-jump or drop abundance profiles (Sch\"oier et
al. 2004). Models of the chemistry of these sources, will then be generated to
explain the origin of these distributions of species and to investigate the
evolution of sources (see Figure 1 of Doty et al. 2004 for a useful graphical explanation of the
entire modeling procedure). The best-fit models will be used to generate and explore
the characteristic, or template, broadband spectra resulting from the
different physical environments in the sources. These spectra will provide a
means of decomposing the broadband 345 GHz spectra of other sources in such a
way as to identify the various physical and chemical processes taking place in
the sources.

\subsection{Chemistry as a Probe of Astrophysics and Evolution}
 
Rarer species are important tracers of specific environments in clouds or
embedded sources.  Examples include the chemistry of deuterium containing
species, sulphur species and complex organic molecules.
 
{\it Deuterium Chemistry}: If fractionation did not take place, a species
containing deuterium would be $> 10^5$ times less abundant than its hydrogen
bearing counterpart and, thus, would be completely undetectable. However, in
cold (T $<$ 20~K) regions, the zero point energy difference between a hydrogen
containing species and its deuterium substituted isotopologue leads to enhanced
abundances of the deuterium containing species (i.e. above the cosmic D to H
ratio).  Recent work on deuterated species have shown that in dense, cold
regions their abundance is enhanced even further, since the species which
would otherwise destroy the deuterated molecules freeze out onto dust grains.  Thus, deuterated molecular species uniquely probe extremely
cold, dense material. For example the deuterated form of a principal ion in
molecular clouds, H$_2$D$^+$, is very weak (0.1~K) towards Class 0 objects
(Stark et al.  1999), but has a 1~K line towards a cold, dense, quiescent
prestellar core where CO is heavily depleted (Caselli et al. 2003).  Clearly
mapping such a line around an embedded protostar will directly trace the
densest, coldest circumstellar material as yet unaffected by the protostar.
 
Conversely, deuterated neutral species are also observed towards several hot cores
 (e.g. Hatchell et al. 1999 and references therein; Pardo et al. 2001;
Osamura et al. 2004). Since the gas temperatures in these regions exceed 50~K
and are often $>$ 100~K (Hatchell et al. 1998a), these species could not have
achieved such high fractionations in their current environment. The most
likely explanation is that they have evaporated from material previously
frozen onto grains as the grains are heated by the protostar (Rodgers \&
Millar 1996). The high deuteration in hot cores points to these grain mantles
having been formed when cold gas, T $<$ 50~K, condensed onto the grains prior
to the formation of, and heating by, an embedded protostar.

{\it Sulphur Species}: The release of species from grain mantles as the grains
are heated injects new species into the gas phase which can be directly
detected, but which also drive new gas-phase reactions. One such reactant
released when grain mantles evaporate is sulphur (Charnley 1997). It may be possible
to use the
subsequent sequential formation of sulphur-bearing daughter species  
 as a chemical clock measuring the time since the mantles were injected,
an idea which has since been investigated for not only high mass sources
(Hatchell et al. 1998b; van der Tak et al. 2003) but also
for lower luminosity sources (Buckle \& Fuller 2003) with more recent
modelling work being done by Wakelam et al. (2004).  Observations,
particularly imaging, of the multiple daughter species in a particular
chemical network is vital in determining the nature of the chemical pathways
involved (J{\o}rgensen et al. 2004). The SLS will be in a unique position to
investigate this issue.
 
{\it Complex Organic Species}: These molecules appear in the spectra of both
low-mass (e.g. Bottinelli et al. 2004; J{\o}rgensen et al. 2005) and high-mass star-forming regions (e.g. Comito et al 2005)
and, in both, may be used as
chemical clocks to indicate source ages and evolutionary status. In prestellar
cores, the organics are very unsaturated and consist of carbon-chains such as
the cyanopolyynes, HC$_{2n+1}$N, n = 1-5, whereas in high-mass regions the
organics are usually highly saturated and very enhanced in abundance, e.g.
methanol, ethanol, C$_2$H$_5$OH, and dimethyl ether, CH$_3$OCH$_3$. In
low-mass regions, the carbon-chains are abundant at early times because the
conversion of carbon to CO has not yet gone to completion. In high-mass
regions, the saturated organics arise as both the direct products (parents) of
grain ice evaporation and as the daughter species which result from a
high-temperature chemistry involving these parents. Nomura and Millar (2004)
have shown how the spatial distributions of molecules in hot core sources can
be used to probe the chemical evolution of the hot gas, the ice composition of
the grains, and the density and temperature profiles of the warm gas. It had
been thought that low-mass star-forming regions had a different chemical
evolution after the formation of a central protostellar source, but the recent
detection of complex saturated organic molecules in IRAS 16293-2422 (Cazaux et
al. 2003) - in a so-called hot corino - has led to a re-evaluation; it is now
thought that these regions may also show the effects of grain evaporation and
subsequent chemical evolution, albeit at somewhat lower temperatures and
densities than in high-mass hot cores.  A sensitive search for organics will
allow us to probe the cloud history, in particular the nature of the material
frozen on ices, the influence of the central protostar in warming its
surrounding natal cloud, the influence of gas temperature on the subsequent
processing of the gas, and the time-scales over which this processing has been
occurring.
 
\subsection{Wider Applications}

The SLS will be a crucial pathfinder experiment for spectroscopic studies of
 star forming regions, by identifying the best combinations
of spectral lines to observe in different types of sources.  Thus, the SLS
sources will serve as templates for observations using next-generation
instruments such as eSMA, ALMA, and Herschel.

However, in addition to providing a basis for understanding the overall
molecular emission from sources in molecular clouds in our galaxy, the results
from the SLS will be important in understanding more distant sources.  An
example of this can be seen in the recent work on NGC253 by Mart\'in et al.
(2005) where observations of sulphur containing species are compared with
galactic regions to determine the state of the molecular clouds in the galaxy.
Another example is the inference that widespread HCO emission in M82 is the
result of the illumination of its molecular clouds by a strong radiation field
(Garc\'{\i}a-Burillo et al. 2002).
 
A further application of complete spectral scans is related to understanding
the continuum emission from sources. For sources with strong spectral lines,
the lines can become important contributors to the broadband flux. For example
towards Orion as much as 60\% of the broadband 345~GHz flux could be due to
lines in the band, not due to continuum emission from dust (White et al. 2003; Groesbeck 1995).
In particular, towards Orion KL the major contaminants are the multitude of
lines of SO$_2$ and SO, CO being only the third most important species
(Schilke et al. 1997).  However, the dominance of SO$_2$ as a contaminant of the
continuum emission is not universal.  
Thompson \& Macdonald (1999) found that in G5.89-0.39, CO was the dominant species, 
possibly due to increased beam dilution for SO$_2$.   The fully sampled maps provided by the
SLS will be able to  determine the spatial distribution of the various molecular species and, thus, identify the underlying cause for differences in the SO$_2$/CO line ratios.

Regardless of the source of contamination,  an accurate determination of the temperature
and density profile from a source requires the measured broadband flux to be
corrected for contamination, something which can only be done with
complete spectral scans.

\section{Target Sources}
 
We will target sources which have been previously studied at other wavelengths
in order to provide continuity with existing studies. All the objects have
also been selected as targets (or potential targets) for the Herschel HIFI
spectral survey.  The SLS observations will consist of a fully sampled
HARP-B footprint  centred on each target source to
determine its molecular inventory.  The desired spectral resolution is 1 km
s$^{-1}$.  This combined spatial and spectral information will, for the first time
in a spectral survey, allow us to direcly determine molecular abundances not
only towards each target, but also in their immediate environment.  
The sources to be observed are described below.

\subsection{Low Mass Source}
  
For the low mass protostars we will observe the well studied source NGC1333
IRAS4. Located in the L1430 core of Perseus, NGC1333 (d $~\sim 320$ pc; de
Zeeuw et al. 1999) is a region forming many YSOs.  IRAS 4 is interesting as it
is one of the first (Mathieu et al. 1994), and one of the youngest (Andr\'e,
Ward-Thompson, \& Barsony 1993), ``proto-binaries'' ever detected.  IRAS 4A
and 4B are separated by a distance of $30''$, and recent evidence has shown
that these sources are binaries themselves.  IRAS 4A has been resolved into
two sources (4A and 4A$'$) separated by 2$''$ (Lay et al.  1995) and IRAS 4B
has been resolved into 4B and 4B$'$ with a 10$''$ separation (Looney et al.
2000).  IRAS 4B may, in fact, be a multiple system itself, with individual
YSOs separated by $~0.5''$.  Both IRAS4A and 4B are associated with molecular
outflows (Blake et al 1995; Choi 2001, 2005).  That from 4A is highly
collimated in the NE-SW direction although the NE lobe shows a sharp bend
interpreted by Choi (2005) as evidence for a jet-core collision.  The outflow
from IRAS 4B, in contrast, is quite compact.  Observations and modeling of
P-cygni profiles in IRAS 4A and 4B has led to YSO masses of 0.71 and 0.23
$M_\odot$ and accretion rates of $1.1\times 10^{-4}$ and $3.7\times 10^{-5}$
$M_\odot$ yr$^{-1}$ respectively (Di Francesco et al. 2001).  The resultant
accretion timescales place the objects at $\sim 6500$ years old.  A number of important molecular species, including some complex molecules (Bottinelli et al. 2004), have been observed at single positions in these two sources (J{\o}rgensen et. al. 2004; 2005; Maret et al. 2004).  These observations
have shown some interesting chemical differences between IRAS 4A \& 4B.  In 4A the 
data are well fit by abundance drops of a factor of 10 to 100 in the dense ($n > 6\times 10^5$ cm$^{-3}$), cold ($T < 40$ K) regions of the envelope.  In contrast, in 4B, the observations can be explained by
an abundance increase in the compact outflow where the temperature rises above 30 K, and not from 
the extended envelope.  HARP-B maps of these to sources will allow us to examine the extended environments of both sources in better detail.

Both IRAS 4A and 4B will be covered in a single, fully sampled HARP-B footprint.  The choice
of 1 km s$^{-1}$ channels is a compromise between a desire to spectrally
resolve lines, covering the frequency band in as few settings as possible and
the required noise level. Although molecular lines from the most quiescent
regions around a low mass protostar may have full width half maxima of less
than 1 km s$^{-1}$, observations of IRAS 16293-2422 indicate that towards this
source a typical line has a width of 1.5-2 km s$^{-1}$ (van Dishoeck et al.
1995; Ceccarelli et al. 2000). The required noise level is set by existing
observations of IRAS 16293 where complex organics tracing the hot gas around
the star are detected at a level of $\sim$ 0.05K, implying the need to reach a
noise level of 0.015 K to detect such lines with a peak signal to noise ratio
of 3. This low noise level will also act to alleviate some of the possible
bandwidth dilution of any narrow lines from the most quiescent gas.
 
\subsection{High Mass Sources}

 For the high mass sources we will observe W49, AFGL2591, and IRAS20126 
 which are believed to span a range of star forming environments and evolutionary states.  
 The lines towards these high-mass objects typically have widths of significantly
 greater than 2 km s$^{-1}$ (Macdonald et al. 1996; Thompson \& Macdonald 1999) and so the
 choice of 1 km s$^{-1}$ resolution was determined by the maximum bandwidth
 with ACSIS. An optically thick line from a 1$''$ diameter hot core with a
 temperature of 100~K towards one of these sources will have a measured line
 peak temperature of 0.4 K in the JCMT beam. To be able to detect the
 optically thin lines which provide most of the information on the sources
 will require reaching noise levels considerably less than this. We intend to
 reach a noise level of 0.04 K per beam in 1 km s$^{-1}$ channels which will be
 smoothed to 2.5 km s$^{-1}$ channels giving a noise level of 0.025 K. This
 noise level is lower than for any of the existing surveys in this band and so
 will probe the chemistry of these regions to a somewhat better depth.

\subsubsection{W49}

W49A was chosen as an example of an extremely active star forming region (e.g. Homeier \& Alves 2005).  Evidence of this comes from the fact that W49A is associated with one of the most powerful H$_2$O masers in the 
Galaxy (Genzel et al. 1978), and that it is also
one of the most luminous regions in the Galaxy 
($L_{bol} \sim 10^7$ L$_\odot$; Ward-Thompson \& Robson 1990).  
W49 is a distant (d = 11.4 kpc; Gwinn et al. 1992) molecular cloud complex consisting of two distinct regions: a supernova remnant (W49B) and a giant HII region (W49A)
separated by $\sim 12'$.  
W49A is composed of a number of optically obscured, 
compact HII regions - W49SW, W49SE and, the strongest,
W49NW (more commonly named W49N; Harvey et al. 1977)
surrounded by a massive molecular cloud ($M > 10^5$ M$_{\odot}$; Mufson \&
Liszt 1977; Simon et al. 2001).
Spectrally, W49A is extremely complex,
containing numerous features contributed by W49A itself, as well as
additional clouds associated 
with the Sagittarius spiral arm, which crosses
the line-of-sight twice.  W49A and the associated line-of-sight clouds
have been the subject of numerous spectroscopic studies (both emission
and absorption) involving species such as H{\sc i} (Radhakrishnan et al. 1972), HCO$^+$ and
HCN (Nyman 1983), OH (Cohen and Willson 1981), H$_2$CO (Mufson and Liszt 1977;
Bieging et al. 1982), CO (Mufson and Liszt 1977), SiO (Lucas and Liszt 2000),
CS (Greaves and Williams 1994), and H$_2$O  (Plume et al. 2004).

\subsubsection{AFGL2591}
 
The bright infrared source AFGL 2591 (L $>$ 20,000 L$_{\odot}$) is a
prototypical object in which to study physical and chemical processes during
high-mass star formation. Located in the
Cygnus-X region, recent observations place it at a distance of 1.7 kpc (Schneider et al 2006). Existing JCMT observations 
(e.g. van der Tak et al. 1999; Doty et al. 2002)
show a rich molecular
line spectrum, which, coupled with its proximity, makes AFGL 2591 a prime
target for the SLS project.  In addition, its physical structure is well
characterized.  The temperature and density structure of the envelope are well
known from the work by Van der Tak et al (1999) and, using this model, the chemical structure
has also been investigated in detail (Doty et al. 2002). A powerful molecular outflow
is visible in high-velocity mid-infrared CO absorption lines, as well as large
lobes of mm-wave emission (Mitchell et al 1990, 1992). High angular resolution
mapping of H$_2$O 22 GHz masers suggests precessing motions in the outflow
(Trinidad et al 2003).  The detection of CO$^+$ and SO$^+$ by St\"auber et al
(2006) suggests that X-ray ionization plays a role for the inner envelope.
Interferometer observations of the H$_2^{18}$O 203 GHz line reveal a flattened
structure with a velocity gradient, possibly a massive circumstellar disk (Van
der Tak et al 2006).

\subsubsection{IRAS~20126+4104}

IRAS20126+4104 is another YSO located in the
Cygnus-X region thus, providing us with two objects which formed at the same distance and, likely, in
similar environments.  Its luminosity is somewhat lower than that of AFGL2591  ($L \sim 10^4 L_\odot$) possibly suggesting a younger evolutionary state.
Recent observations by Cesaroni et al.
(2005) suggest that the YSO mass is $\sim 7 M_\odot$ embedded in a surrounding
0.4 pc diameter core whose mass is $230 M_\odot$ (Estalella et al. 1993).
UKIRT images in K band further suggest that the central object may, in fact,
be a binary system (Sridharan, Williams, \& Fuller 2005).  IRAS20126 has been
known for sometime to contain a powerful molecular outflow oriented NW-SE
which appears to be bounded by an S-shaped distribution of H$_2$ emission
knots (Shepherd et al. 2000; Lebr\'on et al. 2006).  The distribution of the
H$_2$ emission knots have led to the suggestion that the molecular jet in
IRAS20126 is precessing (Shepherd et al. 2000; Cesaroni et al. 2005; Lebr\'on et
al. 2006).  Observations of C$^{34}$S by Cesaroni et al. (2005) have confirmed
the presence of a large Keplerian disk with a radius of 5000 AU.  However,
this disk is thought to be too large to drive the observed outflow.  The
observations of Sridharan et al (2005), however, suggest the presence of a
more compact disk (R $\sim 1000$ AU) which may be more directly related to the
outflow.

\subsection{Photo Dissociation Regions}
  
Based on existing observations and modeling of PDRs we have selected the Orion
Bar region as our target PDR.    This region is subject to a very high UV flux,
$\sim 5\times10^4$ times greater than the typical interstellar field and so is likely
to clearly show the effects of the radiation on the molecular gas.  Previous molecular line observations
(Leurini et al. 2006; Hogerheijde et al. 1995; Jansen et al. 1995) have shown dramatic abundance variations across the bar.  These observations highlight the importance of having combined spatial and spectral information to disentangle complex abundance variations.  Line
widths of 1.8 to 4 km s$^{-1}$ have been observed towards a number of PDR
sources by Schilke et al. (2001). Reaching the same noise level as towards the
high mass protostars would allow use to detect even species as rare as HOC$^+$
at a greater than 4-sigma level (Savage \& Ziurys 2004).

\section{Observing Strategy}

Figure \ref{fig:atmo} shows the atmospheric transmission over the 345 GHz window.
Given the poor transmission at the edges of the window (i.e. below 330 GHz and above 360 GHz)
observations at these frequencies would be prohibitively time consuming.  Thus, the SLS will only cover
330 - 360 GHz.
To cover this bandpass will require 22 tunings of the receiver (although ACSIS will allow a 2GHz bandwidth, the maximum bandwidth per mixer from HARP-B is $\sim$ 1.8GHz) , with
each consecutive frequency tuning overlapping by 10\% of the bandwidth to
avoid gaps in the spectra and to provide consistency between the tunings.
Since HARP-B is a SSB instrument, larger overlaps are not required, since
there is no need to deconvolve emission from different sidebands.  Table \ref{tab:times}
gives the frequency settings and integration times required to achieve uniform rms noise levels.  These values assume grade 4 weather conditions (i.e. 3.75 mm PWV), a source elevation of 90$^\circ$, and a fully sampled HARP-B footprint.

Using ACSIS with a 2 GHz bandwidth will give 0.977 MHz wide channels.  Since for
all the sources except NGC1333 IRAS4, the expected linewidths are greater
than this, for much of the analysis the data will be rebinned to 2.5 \kms\
channels, which is also the channel width used in previous single position
surveys at this frequency band.  The survey
will aim to observe  similar frequency coverage for all the sources.

Table \ref{tab:source} lists the centre positions for the fields for each
sources as well as the  estimated total integration times
required in the Band 4 weather conditions ($0.12 < \tau_{225 GHz}
< 0.2$) allocated to the SLS given the average elevation of the sources.

\subsection{Data Products Provided to the Community}
 
The goal of the SLS is to provide to the astronomical community a uniform and
unbiased spectral survey of a variety of star forming regions.  As such, the
full spectral data-cubes will be publicly available after a one year
proprietary period.  Thus,  we intend to provide:
 
\begin{itemize}
\item Calibrated spectral line datacubes for each observed source.
\item Maps of each molecular species (plus channel maps in the case of
  gradients).
\item Identifications for each emission line seen in the full spectrum of the
  central point in each source.
\item Gas temperatures and column densities for the important species in each
  source.
 \end{itemize}
 
 After the proprietary period has ended, all data products will be available
 from the archive at the Canadian Astronomy Data Centre (CADC\footnote{http://cadcwww.dao.nrc.ca/}).  More details about the SLS and its progress
 will be available on the survey website \footnote{
 http://www.manchester.ac.uk/jodrellbank/research/sls}.

\section{Ancillary Data}

Despite the wealth of information contained in a spectral survey, observations
of single frequency band has a number of limitations.  Clearly there are
species which do not have any transitions in a particular frequency range. As
an example DCO$^+$ and H$_2$D$^+$ do not have any transitions in the 330 - 360 GHz window
covered by the SLS.  There
are also many species, particularly diatomic and linear species, which have
only a single transition in a typical frequency band. Deriving the column
density of such molecules requires more assumptions than for species with
multiple transitions within the band.  In addition there may well be an
excitation mismatch between the transitions of different molecules which are
observed in the band. For example the upper state energy of the CO J=$3-2$
transition, the CO line in the HARP-B band, is 32 K above ground, whereas the
lowest level of any CH$_3$CCH transition in the band is 170 K above
ground. So if there is a range of temperatures along a line of sight these two
species are sensitive to different material.  To overcome these limitations
requires observations at other frequencies. 

The SLS has already started to propose for and collect such observations to
help address these issues in the analysis of the SLS data. For example, we have already obtained
a set of maps covering the HARP-B field of view in a number of species at 230 GHz(See Table
\ref{tab:tab230im}). In addition a full spectral scan from 211 to 279 GHz at a single position has obtained towards AFGL2591.  We have
also been awarded time at the GBT to obtain low frequency spectral scans of the sources covering
the range  8  to 26 GHz. 
This frequency range covers transitions of species including HC$_3$N,
HC$_5$N and HC$_7$N, HDO, C$_4$H, C$_6$H, CCS, H$_2$CO, N$_3$H and the
ground state transitions of CH$_3$CCH and CH$_3$CN.  These GBT
spectral scans will be made at a single position towards the central
source. For comparison with the SLS observations, the HARP-B data will
be convolved down to the GBT beamsize, which over the frequency range
to be observed varies from 84'' to 30''.

\begin{figure}[h] 
   \centering
   \includegraphics[angle=270, width=6in]{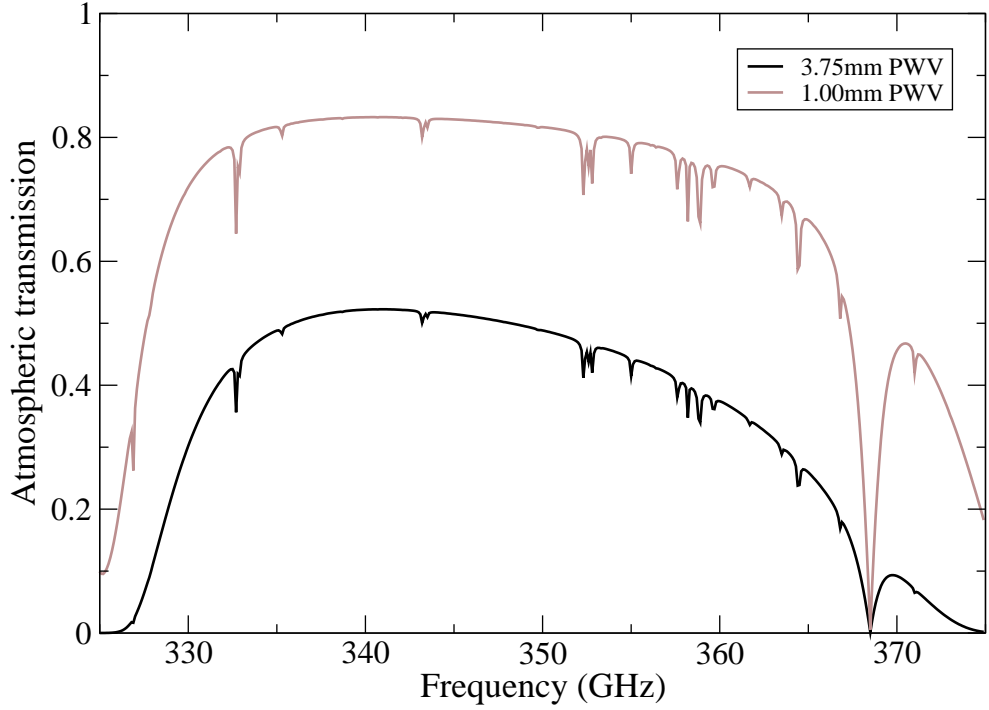} 
   \caption{Atmospheric transmission as a function of frequency for the 345 GHz window.   The upper grey line corresponds to a good night on Mauna Kea (i.e. band 1 weather) with 1mm of precipitable water vapor (PWV) or $\tau_{225 GHz} \approx 0.05$.  The lower black line indicates the band 4 weather in which the SLS will observe, corresponding to 3.75mm PWV or $\tau_{225 GHz} \approx 0.16$.}
   \label{fig:atmo}
\end{figure}

In addition to these ancillary data, we have also been  awarded 76 hours of grade 1-2  weather (i.e 0.5mm $<$ PWV $<$ 1.75mm) at the JCMT
time in 2007.  This time will be used to  observe the SLS sources at frequencies between 360 GHz and 373 GHz which, as Figure \ref{fig:atmo} illustrates, requires good weather conditions.  The 360
- 373 GHz spectral region is the most poorly observed part of the 345 GHz window, but is known to contain
transitions of deuterated species (DCO$^+$, DCN, and H$_2$D$^+$), formaldehyde, and methanol together
with sulphur bearing species, which can probe different evolutionary stages of the star formation process.

These data, together with any other data taken by the SLS collaboration, will
also be made publically available.

\newpage

\newpage
\begin{table}[h]
\caption{SLS Frequencies \& Integration Times in hours (assuming 90$^\circ$ source elevation)}
\begin{center}
\begin{tabular}{lcc}
\hline\hline
Frequency & Low-Mass Sources & High Mass Sources \\
(GHz) & $T_{rms} = 0.03 ~^1$ K & $T_{rms} = 0.08 ~^1$ K \\
\hline \hline
 332.237 & 6.62 & 0.93 \\
 333.553 & 5.51 & 0.77 \\
 334.868 & 4.94 & 0.69 \\
 336.184 & 4.61 & 0.65 \\
 337.500 & 4.40 & 0.62 \\
 338.816 & 4.29 & 0.60 \\
 340.132 & 4.22 & 0.59 \\
 341.447 & 4.20 & 0.59 \\
 342.763 & 4.22 & 0.59 \\
 344.079 & 4.25 & 0.60 \\
 345.395 & 4.30 & 0.60 \\
 346.711 & 4.38 & 0.62 \\
 348.026 & 4.48 & 0.63 \\
 349.342 & 4.61 & 0.65 \\
 350.658 & 4.77 & 0.67 \\
 351.974 & 5.04 & 0.71 \\
 353.289 & 5.25 & 0.74 \\
 354.605 & 5.50 & 0.77 \\
 355.921 & 5.83 & 0.82 \\
 357.237 & 6.33 & 0.89 \\
 358.553 & 7.14 & 1.00 \\
 359.868 & 7.77 & 1.09 \\
 
  \hline \hline
  $^1$ - assuming a velocity resolution of 2.5 km s$^-1$
\end{tabular}
\end{center}
\label{tab:times}
\end{table}

\newpage
\begin{table}[h]
\caption{SLS Source List \& Total Integration Times}
\begin{center}
\begin{tabular}{lcccccc}
\hline\hline
Source & RA(J2000) & DEC (J2000)& Avg. El. & Total Int. Time\\
              & (hh:mm:ss.s) & (dd:$'$:$''$) &  & (hrs) \\
\hline \hline
 
 NGC1333 IRAS 4 & 03:29:11.1 & +31:13:40 &  60$^\circ$ & 115.6 \\
 W49                         & 19:10:13.4 & +09:06:14 & 60$^\circ$ & 16.8 \\
 AFGL2591              & 20:29:24.9 & +40:11:20 &  60$^\circ$ & 16.8 \\
 IRAS20126+4104 & 20:14:26.0 & +41:13:32 &  60$^\circ$ & 16.8 \\
 Orion Bar                & 05:35:22.5 & -05:25:00  &  50$^\circ$ & 18.8 \\
  \hline \hline
{\bf Total Time (hrs)}&                  &                   &             &         {\bf 184.8} \\
\end{tabular}
\end{center}
\label{tab:source}
\end{table}

\begin{table}
  \centering
   \caption{Species detected and imaged by SLS Survey at  230 GHz.}
  \begin{tabular}{ll}
Source & Species \\ \hline
          W49 & $^{13}$CO, C$^{18}$O, C$^{17}$O, HCN, H$^{13}$CN, H$^{13}$CO$^+$, 
CH$_3$OH, SiO\\
     OrionBar & $^{13}$CO, C$^{18}$O, C$^{17}$O,  HCN\\
NGC1333~IRAS4 & $^{13}$CO, C$^{18}$O, C$^{17}$O, HCN, H$^{13}$CO$^+$, CH$_3$OH, DCO$^+$,\\
     AFGL2591 & $^{13}$CO, C$^{18}$O, HCN, H$^{13}$CN, H$^{13}$CO$^+$, CH$_3$OH\\ 
    IRAS20126 & $^{13}$CO, C$^{18}$O, C$^{17}$O, DCN, CH$_3$OH, H$_2$S(?)   \\
\hline \hline
  \end{tabular}
  \label{tab:tab230im}
\end{table}

 \end{document}